\documentclass[aps,pre,twocolumn,tight,superscriptaddress,showpacs]{revtex4}
\usepackage{amssymb}
\usepackage{color}
\usepackage{graphicx}
\usepackage{amsmath}
\usepackage{times}
\usepackage{bm}

\begin{document}

\title{Evoking complex neuronal networks by stimulating a single neuron}

\author{Mengjiao Chen}
\affiliation{School of Physics and Information Technology, Shaanxi Normal University, Xi'an 710062, China}
\author{Weijie Lin}
\affiliation{School of Physics and Information Technology, Shaanxi Normal University, Xi'an 710062, China}
\affiliation{Department of Physics, Zhejiang University, Hangzhou 310027, China}
\author{Hengtong Wang}
\affiliation{School of Physics and Information Technology, Shaanxi Normal University, Xi'an 710062, China}
\author{Wei Ren}
\affiliation{College of Life Science, Shaanxi Normal University, Xi'an 710062, China}
\affiliation{Key Laboratory of MOE for Modern Teaching Technology, Shaanxi Normal University, Xi'an 710062, China}
\author{Xingang Wang}
\email[Corresponding author. Email address: ]{wangxg@snnu.edu.cn}
\affiliation{School of Physics and Information Technology, Shaanxi Normal University, Xi'an 710062, China}
\begin{abstract}

The dynamical responses of complex neuronal networks to external stimulus injected on a \emph{single} neuron are investigated. Stimulating the largest-degree neuron in the network, it is found that as the intensity of the stimulus increases, the network will be transiting from the resting to firing states and then restoring to the resting state, showing a bounded firing region in the parameter space. Furthermore, it is found that as the coupling strength decreases, the firing region is gradually expanded and, at the weak couplings, separated into disconnected subregions. By a simplified network model, we conduct a detail analysis on the bifurcation diagram of the network dynamics in the two-dimensional parameter space spanned by stimulating intensity and coupling strength, and, by introducing a new coefficient named effective stimulus, explore the mechanisms of the modified firing region. It is revealed that the coupling strength and stimulating intensity are equally important in evoking the network, but with different mechanisms. Specifically, the effective stimuli are \emph{shifted up} globally with the increase of the stimulating intensity, while are \emph{drawn closer} with the increase of the coupling strength. The dynamical responses of small-world and random complex networks to external stimulus injected on the largest-degree neuron are also investigated, which confirm the generality of the observed phenomena.

\end{abstract}

\date{\today}
\maketitle

\section{Introduction}

The stimulus-response relationship of neuronal systems is of focusing interest in neuroscience \cite{Neuroscience}. Whereas the dynamical responses of a single neuron to various kinds of stimuli have been well explored and documented \cite{Izhikevich2007}, the stimulus-response relationship for complex neuronal systems made up of an ensemble of coupled neurons remains an open question \cite{BrainStimulation}. This is particularly the case for the excitability of coupled neuronal systems when subjected to external stimulating currents, where the bifurcation diagram of a single neuron has been well explored in both experiment and theory, yet the corresponding picture for complex neuronal systems is still not clear. For a single neuron, as the intensity of the stimulating current increases, it is widely observed that the neuron dynamics typically will undergo two bifurcations: the transition from the polarized-resting to firing states (at a smaller critical intensity named firing threshold), and the transition from the firing to depolarized-resting states (at a larger critical intensity named depolarization-block threshold) \cite{Izhikevich2007}. However, for the coupled neuronal systems, e.g., the complex neuronal networks, despite the experimental efforts made in the past decade, it remain unknown whether such transitions still exist and, more importantly, how the transitions are influenced by the neuron couplings \cite{Brecht 2004,Bi 2005,Houweling 2008,Morgan 2008,C. Y. Li 2009}.

The excitability of complex neuronal systems is fundamentally different from that of isolated neuron, due to the neuron couplings \cite{Houweling 2008}. When coupled together, the excitability of a neuron is jointed determined by the external stimulus and the coupling signals it receives from the connected neurons. As such, the bifurcation thresholds mentioned above, namely the firing and depolarization-block thresholds, will be \emph{dependent} on also the strength of the neuron couplings \cite{Fontanini 2008}. Meanwhile, adapting to the neuronal activities, the neuron coupling strength could be varying with time, i.e., the plasticity feature of the neuron synapses (junctions) \cite{Kothmann 2012,Landisman 2005,Connors 2004}. Therefore, even for the same neuronal system, the bifurcation thresholds might be different when measured under different physiological conditions. Besides the aspect of coupling strength, the excitability of complex neuronal systems is also crucially affected by the coupling relationship (structure) among the neurons. Modern studies of brain connectome have revealed that neuronal systems are more properly described by complex networks \cite{BrainNet}, where neurons and synapses are represented as nodes and links, respectively. Instead of random connections, the neuronal networks are well structured and possess distinct topological properties, e.g., the heterogenous degree distribution \cite{Egu2005,Sporns2007}, the community and modular structures \cite{Milo 2002,Reigl 2004,Sporns 2004}, the hierarchical organization \cite{Felleman 1991,Hilgetag 2000}, etc. In particular, in neuronal networks there are a few neurons of very large degrees (the number of connections of a neuron), which, comparing to the majority of neurons having smaller degrees, are more crucial to the system functionalities \cite{Egu2005,Sporns2007}. As such, the excitability of neuronal networks will be also affected by the location (node) where the stimulus is added.

Experiments aside, insights into the excitability of complex neuronal networks could be firstly gained by modeling studies. Inspired by this, in the present work we investigate theoretically how complex network of coupled neurons is responded to external stimulus, with special attentions being paid to the dependence of network excitability on the stimulating intensity and neuron coupling strength. More specifically, we stimulate only a single node, i.e., the largest-degree node, in a heterogenous complex network, and investigate how the network firing thresholds are varying with the stimulating intensity and neuron coupling strength. Our main finding is that, by stimulating just a single node, the whole network can still be successfully evoked when the stimulating intensity is within a bounded region in the parameter space (the interval between the firing and depolarization-block thresholds); and, with the variation of the neuron coupling strength, the range of this firing region could be significantly modified, e.g., the firing region is separated into disconnected subregions under weak couplings. Our studies show clearly the existence of both two firing thresholds in complex neuronal networks and, more importantly, point out the dependence of the firing thresholds on the neuron couplings.

\section{Model}

Our model of networked neurons reads
\begin{equation}
\dot{\mathbf{x}}_i=\mathbf{F}(\mathbf{x}_i,I_i)+\varepsilon\sum\limits^{N}_{j=1}a_{ij}\mathbf{H}(\mathbf{x}_j),
\end{equation}
with $i,j=1,2,\ldots,N$ the neuron (node) indices, $\mathbf{x}_i$ the state vector of the $i$th neuron, and $\varepsilon$ the uniform coupling strength. $\dot{\mathbf{x}}_i=\mathbf{F}(\mathbf{x}_i,I_i)$ describes the local dynamics of the $i$th neuron, with $I_i$ the bifurcation parameter. $\mathbf{H}(\mathbf{x})$ denotes the way how the neurons are coupled with each other, i.e., the coupling function. The coupling relationship among the neurons is captured by the adjacency matrix $\mathbf{A}=\{a_{ij}\}$, with $a_{ij}=a_{ji}=1$ if nodes $i$ and $j$ are directly connected, and $a_{ij}=0$ otherwise. The diagonal elements of $\mathbf{A}$ are set as $a_{ii}=-\sum_j a_{ij}=-k_i$ (the degree of node $i$), so as to capture the diffusive coupling between neurons \cite{Buono 2001}.

In our studies, we adopt the two-dimensional Morris-Lecar (ML) model to characterize the neuron dynamics, which in its isolated form is described by the equations \cite{Morris 1981}
\begin{eqnarray}
C\dot{V} &=& -g_{Ca}m_{\infty}(V)(V-V_{Ca})\nonumber \\
          & & -g_Kn(V-V_K) -g_L(V-V_L)+I, \\
\dot{n} &=& \phi [n_{\infty}(V)-n]/\tau_{n}(V),
\end{eqnarray}
where
\begin{eqnarray}
m_\infty(V)&=&\frac{1}{2}[1+\tanh(\frac{V-a_1}{a_2})], \nonumber \\
n_\infty(V)&=&\frac{1}{2}[1+\tanh(\frac{V-a_3}{a_4})], \nonumber \\
\tau_{n}(V)&=&1/{\cosh (\frac{V-a_3}{2a_4})}\nonumber.
\end{eqnarray}
Here, $V(t)$ and $n(t)$ represent, respectively, the potential and channel activity (the opening probability of the potassium ion channels) of the neuron membrane. In Eq. (2), $C$ denotes the membrane capacitance; $g_{Ca}$, $g_K$, and $g_L$ denote, respectively, the maximal conductances of the calcium ($Ca^{2+}$), potassium ($K^{+}$), and leakage ion currents; $V_{Ca}$, $V_K$, and $V_L$ denote the equilibrium potentials of the corresponding ion channels; $m_\infty(V)$ denotes the opening probability of the calcium conductance in equilibrium; and $I$ denotes the intensity of the externally added current. In Eq. (3), $\phi$ denotes the reference frequency, $n_{\infty}(V)$ denotes the opening probability of the potassium conductance in equilibrium, and $\tau_{n}(V)$ characterizes the time scale of the potassium channel. The ML model is originally proposed for reproducing the variety of oscillatory behaviors in relation to calcium and potassium conductance in the giant barnacle muscle fiber \cite{Morris 1981}, and, as the simplified version of the four-dimensional Hodgkin-Huxley model \cite{Hodgkin 1948}, has been widely employed in literature in exploring the firing behaviors of various neuronal systems \cite{Gutkin 1998,Agnon-Snir 1998,Buono 2001,Sejnowski 2013}.

Following Ref. \cite{Tsumoto 2006}, in the ML model we set the parameters $(g_{Ca},g_{K},g_{L})=(4.4~ms/cm^2,8.0~ms/cm^2,2.0~ms/cm^2)$, $(V_{Ca},V_K,V_L)=(120~mV,-84~mV,-60~mV)$, $\phi=0.04$, $C=20.0~\mu F/cm^2$, $a_1=-1.2~mV$, $a_2 =18~mV$, $a_3 =2~mV$, and $a_4=30~mV$. With such a setting, a single neuron will be staying on the firing states when the intensity of the stimulating current is within the range $I\in(I^s_1,I^s_2)$, with $I^s_1\approx 93.86~pA$ (the firing threshold) and $I^s_2\approx 212.02~pA$ (the depolarization-block threshold); while for $I\leq I^s_1$ or $I\geq I^s_2$, the neuron will be staying on the resting states \cite{Izhikevich2007}. The key question we are going to address in the present work is: \emph{Is such a firing region still existing for complex network of coupled neurons? and, if yes, how is this firing range affected by the neuron coupling strength?}

To capture the heterogenous feature of the neuronal networks, we adopt the generalized Barab\'asi-Albert (BA) algorithm to construct the network structure \cite{Barabasi 1999}. Specifically, starting from a small-size nucleus of $7$ globally connected nodes, at each time step of the network growth a new node is introduced and is connected to $2$ of the existing nodes by the probability $\prod_i=k_i/\sum_j k_j$, with $i$ and $j$ the node indices and $k_i$ the degree of node $i$ (i.e., the mechanism of preferential link attachment). The network growth is stopped at $N=50$, with the largest hub has the degree $k_{max}=20$. According to the network links, we couple the neurons with the function $\mathbf{H}([V,n]^T)=[V,0]^T$ (i.e., the diffusive coupling of the membrane potentials \cite{Bennett 2004}) by the uniform strength $\varepsilon$ (which has the unit $pA/mV$). The initial conditions of the neurons are set uniformly as $(V_i,n_i)=(-10,0)$, and Eq. (1) is numerically solved by the 4th-order Runge-Kutta method with the time step $\delta t=1\times 10^{-2} ~ms$. In the absence of stimuli, the network will settle to the resting state after a short transient. In our studies, we shall stimulate only the largest-degree node by injecting on it a direct current of intensity $I$, and analyze the bifurcation diagram of the network dynamics in the two-dimensional parameter space $(I,\varepsilon)$.

\section{Numerical results}

\begin{figure*}[tbp]
\begin{center}
\includegraphics[width=0.8\linewidth]{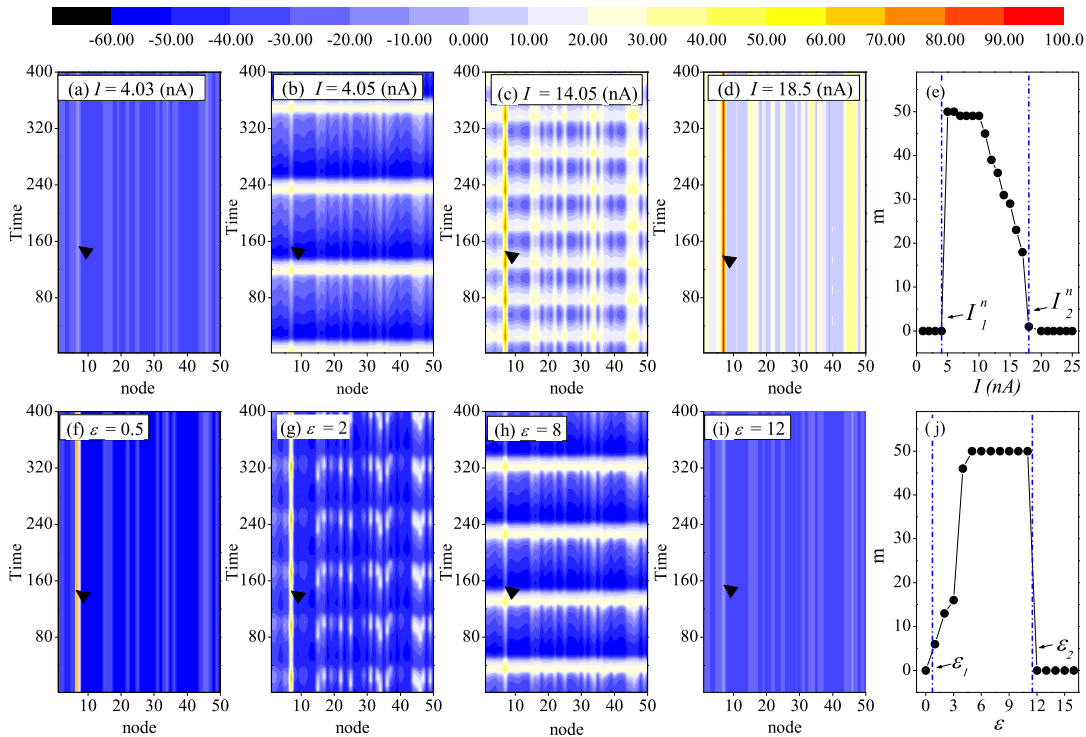}
\caption{(Color online) The variations of the network dynamics as functions of the stimulating intensity $I$ (a-e) and the coupling strength $\varepsilon$ (f-j). Fixing the coupling strength as $\varepsilon=12$, the spatiotemporal evolutions of the neuron membrane potentials for $I=4.03~nA$ (a), $I=4.05~nA$ (b), $I=14.05~nA$ (c), and $I=18.5~nA$ (d). (e) The variation of the number of firing neurons, $m$, as a function of $I$. The network is evoked in the region $(I^n_1,I^n_2)$, with $I^n_1\approx 4.04~nA$ and $I^n_2\approx 18.0~nA$. Fixing the stimulating intensity as $I=4~nA$, the spatiotemporal evolutions of the network for $\varepsilon=0.5$ (f), $\varepsilon=2$ (g), $\varepsilon=8$ (h), and $\varepsilon=12$ (i). (j) $m$ versus $\varepsilon$. The network is evoked in the region $(\varepsilon_1,\varepsilon_2)$, with $\varepsilon_1\approx 0.7$ and $\varepsilon_2\approx 11.3$. Arrows denote the location of the largest-degree node where the stimulus is injected.}
\label{Fig.1}
\end{center}
\end{figure*}

We start by exploring the variation of the network excitability with respect to the stimulating intensity, while keeping the coupling strength fixed as $\varepsilon=12$. Setting $I=4.03~nA$, we plot in Fig. 1(a) the spatiotemporal evolution of the network after a transient period of $T=2.6\times 10^3~ms$. It is seen that all the neurons, including the largest-degree one, are of steady membrane potentials, indicating that the network is completely resting. Increasing $I$ to $4.05~nA$, in Fig. 1(b) it is seen that all the neurons are firing in a synchronous fashion, i.e., the whole network is evoked. We thus infer from Figs. 1 (a) and (b) that, despite the high-dimensional dynamics, complex neuronal network can still be evoked by stimulating a single neuron; and, in analogy with the excitability of a single neuron, there also exists a firing threshold beyond which the network can be evoked. This triggers our interest of finding the depolarization-block threshold by increasing $I$ further. Fig. 1(c) shows the network evolution for $I=14.05~nA$. It is seen that the network is partially evoked, i.e., some of the neurons are ceased from oscillations, while the others remain firing. Increasing $I$ further to $18.5~nA$ [Fig. 1(d)], it is shown that all the neurons are ceased, and the network restores to the completely resting state similar to Fig. 1(a). Indeed, the depolarization-block threshold is observed. Here, to distinguish from the situation of isolated neuron, we use $I^n_1$ and $I^n_2$ to denote, respectively, the firing and depolarization-block thresholds of the neuronal network.

To have more details on the transition of the network dynamics with respect to $I$, we plot in Fig. 1(e) the number of firing neurons, $m$, as a function of $I$. Here, neurons are identified as firing if their membrane potentials are oscillating with amplitudes larger than $20~mV$ (so as to exclude the situations of subcritical and passive oscillations). As depicted in Fig. 1(e), as $I$ increases from 0, the value of $m$ is firstly jumped from 0 (resting network) to $50$ (completely firing network) at $I=I^n_1\approx 4.04~nA$; then, during the interval $(I^n_1,6.95~nA)$, $m$ is staying at $50$; after that, $m$ is gradually decreased, and reaches 0 at $I=I^n_2\approx 18.0~nA$. The transition scenario of the network dynamics therefore can be described as: resting $\rightarrow$ complete-firing $\rightarrow$ partial-firing $\rightarrow$ resting, and the firing region of the network (including both the complete- and partial-firing states) is identified as $(I^n_1,I^n_2)=(4.04~nA,18.0~nA)$.

We continue to explore the influence of the neuron coupling strength, $\varepsilon$, on the network excitability. Fixing the stimulating intensity as $I=4~nA$, we plot in Figs. 1(f-i) the network evolutions under different coupling strengths. It is straightforward to see that when the coupling is absent ($\varepsilon=0$), the network will be at resting, since $I>I^s_2$ for the largest-degree neuron and $I=0<I^s_1$ for the others. By the weak coupling $\varepsilon=0.5$, Fig. 1(f) shows that the network is still at resting; increasing $\varepsilon$ to $2$, the network is partially evoked [Fig. 1(g)]; when $\varepsilon=8$, the network is completely evoked [Fig. 1(h)]; increasing $\varepsilon$ further to $12$, the network restores to the resting state [Fig. 1(i)]. Again, we observe the transition from the resting to firing and then to the resting states. More details about the transition can be found in Fig. 1(j), where the number of the firing neurons, $m$, is plotted as a function of $\varepsilon$. Fig. 1(j) shows that the network is evoked (partially or completely) in the region $(\varepsilon_1,\varepsilon_2)$, with $\varepsilon_1\approx 0.7$ and $\varepsilon_2\approx 11.3$. Clearly, the network excitability is also influenced by the neuron coupling strength.

\begin{figure}[tbp]
\begin{center}
\includegraphics[width=0.78\linewidth]{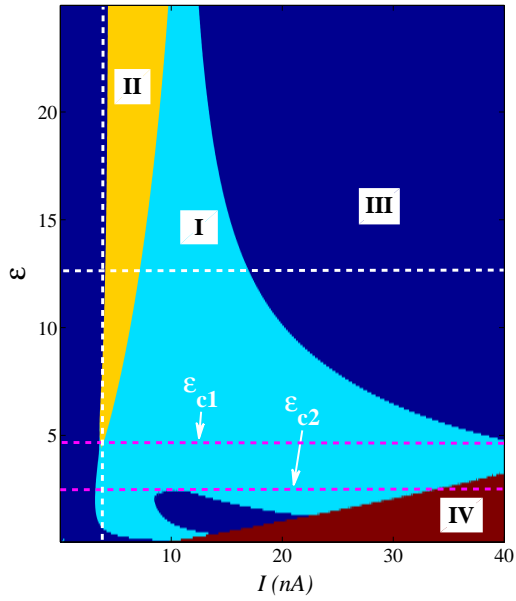}
\caption{(Color online) Bifurcation diagram of the network dynamics in the two-dimensional parameter space $(I,\varepsilon)$. As $\varepsilon$ increases from $\varepsilon_{c1}\approx 4.9$, the firing (depolarization-block) threshold, $I^n_1$ ($I^n_2$), is monotonically increased (decreased), rendering the firing region (regions I and II) being gradually narrowed. For $\varepsilon<\varepsilon_{c2}\approx 2.45$, the firing region is separated into two branches by an embedded resting ``tongue". Regions I, II, III, and IV represent, respectively, the partial-firing, complete-firing, resting, and overflowing states. The (white) horizontal (vertical) dashed line denotes the transition process presented in Figs. 1(a-e) [Figs. 1(f-j)]. The (magenta) dotted lines denote the two critical coupling strengths $\varepsilon_{c1}$ and $\varepsilon_{c2}$.}
\label{Fig.2}
\end{center}
\end{figure}

To have a global picture on the bifurcation of the network dynamics, we go on to scan the network states over a certain region in the two-dimensional parameter space $(I,\varepsilon)$. The results are presented in Fig. 2. Now, the influence of the neuron coupling strength on the network excitability can be systematically analyzed. When the coupling is strong ($\varepsilon>\varepsilon_{c1}\approx 4.9$), it is seen that with the increase of $\varepsilon$, the firing (depolarization-block) threshold, $I^n_1$ ($I^n_2$), is monotonically increased (decreased). As a consequence, the firing region is gradually narrowed with the increase of $\varepsilon$. A close look to the firing region also shows that with the increase of $\varepsilon$, the region of complete firing (region II in Fig. 2) is gradually expanded. When the coupling is weak ($\varepsilon<\varepsilon_{c2}\approx 2.45$), the variation of the firing region is significantly different from that of strong couplings, and shows some intriguing features. Firstly, the firing region is separated into two disconnected subregions by an embedded ``tongue" of resting states. As such, with the increase of $I$, the network will be transiting twice from the resting to firing states, instead of once as for the case of strong couplings. Secondly, in contrast to the situation of strong couplings where $I^n_1$ is monotonically increased with $\varepsilon$, here it is seen that $I^n_1$ is monotonically decreased as $\varepsilon$ increases. Finally, an overflowing region is appeared in the bottom-right corner of the parameter space (region IV in Fig. 2). As a result of this, the network will be transited from the firing to overflowing states at large $I$, instead of restoring to the resting state as for the case of strong couplings.

\section{Mechanism Analysis}

How could complex neuronal network be evoked by stimulated only a single neuron, and why the firing region is modulated by the coupling strength in such a fashion? In particular, why the firing region is separated into disconnected subregions at weak couplings, while is continuously distributed at strong couplings? To answer these questions, we next employ a simplified model to explore the underlying mechanisms of the modified network excitability induced by varying the neuron coupling strength.

\begin{figure}[tbp]
\begin{center}
\includegraphics[width=\linewidth]{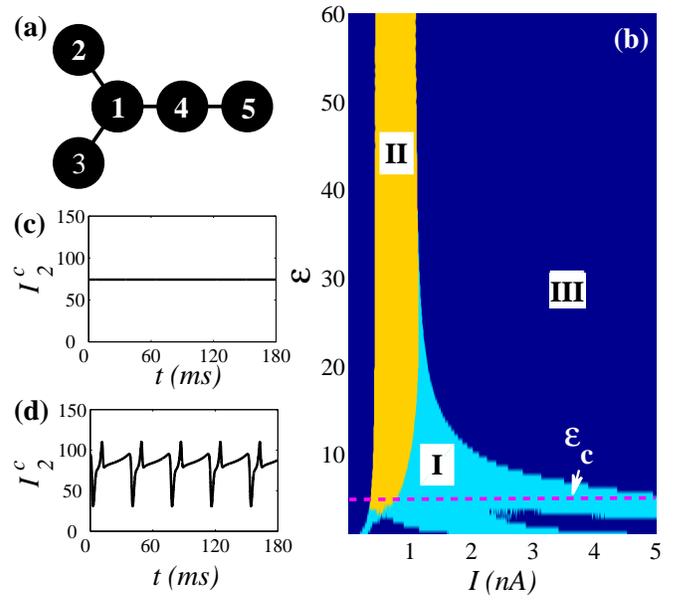}
\caption{(Color online) The excitability of a simplified neuronal network. (a) The network structure. Stimulus is injected on the $1$st node which has degree $k_1=3$. (b) The bifurcation diagram in the parameter space $(I,\varepsilon)$. Regions I, II, and III represent, respectively, the partial-firing, complete-firing, and resting states. The firing region is separated into three disconnected subregions in the region of $\varepsilon<\varepsilon_c\approx 3.2$ (the magenta dotted line). Fixing the stimulating intensity as $I=420~pA$, the time evolution of the coupling signal that neuron 2 receives from neuron 1, $I^c_2$, under different coupling strengths: (c) $\varepsilon=2$ (steady); (d) $\varepsilon=8$ (oscillatory).}
\label{Fig.3}
\end{center}
\end{figure}

The new network structure is presented in Fig. 3(a), which contains $5$ nodes and $4$ links. Stimulating the $1$st neuron (the largest-degree node), we investigate again the bifurcation diagram of the network dynamics in the parameter space $(I,\varepsilon)$. The results are presented in Fig. 3(b). It is seen that, despite the simplified network model, Fig. 3(b) reproduces the main features shown in Fig. 2. To be specific, the firing region is gradually narrowed as $\varepsilon$ increases from $\varepsilon_c\approx 3.2$, while is separated into three disconnected subregions for $\varepsilon<\varepsilon_c$. The similarity between Figs. 2 and 3 renders the simplified model a suitable candidate for exploring the dynamical mechanisms of the modified network excitability.

As only the largest-degree neuron is stimulated, the other neurons in the network therefore are evoked by their nearest neighbors through the couplings. Taking neuron 2 in Fig. 3(a) as an example, the coupling signal that it receives from neuron 1 is $I^c_2=\varepsilon(V_1-V_2)$ (according to the coupling function). Assuming the network is at resting, $I^c_2$ will be a constant, which is essentially a direct current as the one injected on neuron $1$. Fig. 3(c) shows the case for $I=420~pA$ and $\varepsilon=2$, where $I^c_2$ is fixed at $74.3~pA$ during the system evolution. As $\varepsilon$ increases, the coupling current will be gradually increased and, once $I^c_2$ exceeds $I^s_1$ (the firing threshold for isolated neuron), neuron 2 could be fired. In this case, $I^c_2$ will be oscillating with time, as depicted in Fig. 3(d) for the case of $I=420~pA$ and $\varepsilon=8$. Neuron $1$, in turn, might be evoked by the feedback coupling, $-I^c_2$, given the total current it receives [including the stimulus $I$, the feedback couplings $-I^c_2$ (from neuron 2) and $-I^c_3$ (from neuron 3)] is within the firing region $(I^s_1,I^s_2)$. In the similar way, the other neurons in the network could be also evoked by the coupling currents, resulting in the complete firing of the network.

\begin{figure*}[tbp]
\begin{center}
\includegraphics[width=0.8\linewidth]{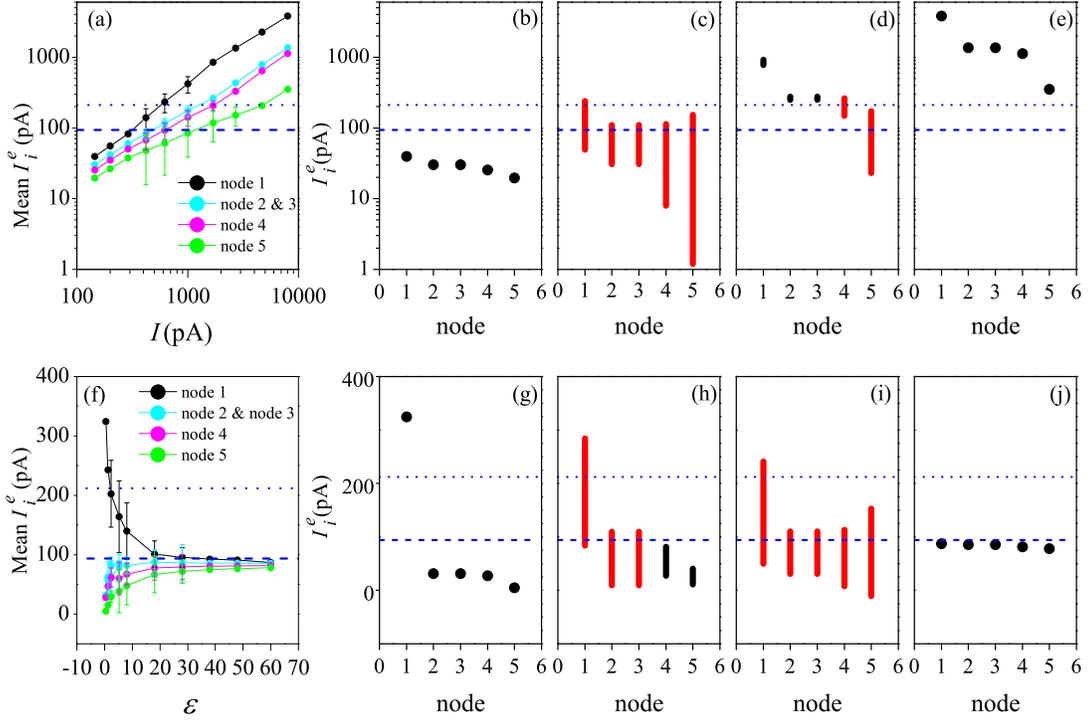}
\caption{(Color online) For the simplified network model shown in Fig. 3(a), the variations of the effective stimuli, $I^e_i$, with respect to the stimulating intensity $I$ (upper panels) and the coupling strength $\varepsilon$ (lower panels). Fixing $\varepsilon=8$, the variations of $I^e_i$ as a function $I$ (a), and the distribution of $I^e_i$ for different stimulating intensities: (b) $I=145~pA$ (resting), (c) $I=420~pA$ (complete firing), (d) $I=1.7~nA$ (partial firing), and (e) $I=8~nA$ (resting). Fixing $I=420~pA$, the variations of $I^e_i$ as a function $\varepsilon$ (f), and the distribution of $I^e_i$ for different coupling strengths: (g) $\varepsilon=0.5$ (resting), (h) $\varepsilon=2.4$ (partial firing), (i) $\varepsilon=8$ (complete firing), and (j) $\varepsilon=60$ (resting). Dashed and dotted horizontal lines represent, respectively, the firing and depolarization-block thresholds of isolated neuron. In (b-e) and (g-j), the error bars denote the oscillating amplitudes of $I^e_i$, and are red-colored for the firing neurons.}
\label{Fig.4}
\end{center}
\end{figure*}

The above analysis thus suggests that in evoking the networked neurons, the coupling currents, $I^c$, play essentially the same role as the stimulating current, $I$. Regarding this, we unify the coupling and stimulating currents by introducing the new quantity of effective stimulus
\begin{equation}
I^e_i=I_i+\varepsilon\sum^N_{j=1}a_{ij}(V_j-V_i).
\end{equation}
If $I^e_i$ is oscillating with time, we characterize it by its time-average $\left<I^e_i\right>$ and oscillating amplitude $A$. Now, whether neuron $i$ is resting or firing is solely determined by the integrated signal $I^e_i$, regardless of the details of the stimulus (e.g., the stimulating intensity) and couplings (e.g., the set of neighboring nodes). Moreover, with the introduction of effective stimulus, the complex network can be treated effectively as an ensemble of \emph{isolated} neurons subjected to different effective stimuli. As we shall show in the following, this technique of neuron decoupling is very favorable in analyzing the firing mechanisms of complex neuronal networks.

With the help of effective stimulus, we now revisit, from a different viewpoint, the influences of $I$ and $\varepsilon$ on the network excitability. Fixing $\varepsilon=8$, we plot in Fig. 4(a) the variations of $I^e_i$ as a function of $I$ for all 5 neurons in the network. Interestingly, it is seen that with the increase of $I$, the effective stimuli are \emph{shifted up} globally. Specifically, the effective stimuli are monotonically increased and crossing the firing region $(I^s_1,I^s_2)$ in sequence ($I^e_1>I^e_2=I^e_3>I^e_4>I^e_5$). Another interesting phenomenon observed in Fig. 4(a) is that a neuron is firing only when its effective stimulus is overlapping with the firing region. To have more details on the connection between the network dynamics and the effective stimuli, we plot in Figs. 4(b-e) the distributions of $I^e_i$ for some typical values of $I$. For $I=145~pA$ [Fig. 4(b)], the effective stimuli are constants and below the firing threshold $I^s_1$, while a check of the neuron dynamics shows that the network is at resting. For $I=420~pA$ [Fig. 4(c)], the effective stimuli are oscillating and overlapping with the firing region. By checking the neuron dynamics, it is found that the network is completely fired. Increasing $I$ to $1.7~nA$ [Fig. 4(d)], $I^e_{1,2,3}$ are larger than the depolarization-block threshold $I^s_2$, while $I^e_{4,5}$ are still overlapping with the firing region. In this case, only neurons $4$ and $5$ are fired, and, as a consequence, the network is partially evoked. Increasing $I$ further to $8~nA$ [Fig. 4(e)], all the effective stimuli become constants and are larger than $I^s_2$, and the network is restored to the resting state similar to Fig. 4(b). The role of $I$ in evoking the network now can be understood: it increases the effective stimuli of all the neurons, making them crossing the firing region, $(I^s_1,I^s_2)$, in sequence.

We finally analyze the influence of the coupling strength on network excitability. Fixing $I=420~pA$, we plot in Fig. 4(f) the variations of the effective stimuli, $I^e_i$, as a function of $\varepsilon$. Different from the situation of increasing $I$ [Fig. 4(a)], it is seen that as $\varepsilon$ increases, the effective stimuli are \emph{drawn closer}. To be specific, with the increase of $\varepsilon$, $I^e_1$ is quickly decreased from above $I^s_2$ to below $I^s_1$, while, in the meantime, $I^e_{2,3,4,5}$ are slowly increased and approaching $I^s_1$. To have more details about the transition, we plot in Figs. 4(g-j) the distribution of $I^e_i$ for different values of $\varepsilon$. For the weak coupling $\varepsilon=0.5$ [Fig. 4(g)], a distinct gap is observed between $I^e_1$ and $I^e_{2,3,4,5}$, with $I^e_1>I^s_2$ and $I^e_{2,3,4,5}<I^s_1$. In this case, as all the effective stimuli are outside of the firing region, no neuron is firing and the network is completely resting. Increasing $\varepsilon$ to $2.4$ [Fig. 4(h)], $I^e_1$ is decreased while $I^e_{2,3,4,5}$ are increased, making the gap between them narrowed. Moreover, the oscillations of $I^e_{1,2,3}$ are overlapping with the firing region, while $I^e_{4,5}$ are not. In this case, neurons $1$, $2$ and $3$ are firing, making the network partially fired. Increasing $\varepsilon$ to $8$ [Fig. 4(i)], the gap between $I^e_1$ and $I^e_{2,3,4,5}$ is further narrowed, and all the effective stimuli are overlapping with the firing region. In this case, the network is completely evoked. Increasing $\varepsilon$ further to $60$ [Fig. 4(j)], the effective stimuli have approximately the same value, i.e., $I^e_1\approx I^e_{2,3,4,5}$, and all are smaller to $I^s_1$. In this case, all neurons are ceased from oscillations and the network is restored to the resting state. The role of $\varepsilon$ in evoking the neuronal network thus can be also understood: it reduces the difference between the effective stimuli (especially the gap between the stimulated and non-stimulated neurons), drawing them to the uniform distribution.

Having revealed the individual influence of $I$ and $\varepsilon$ on the network excitability, the bifurcation diagrams shown in Figs. 2 and 3(b) now can be interpreted, as follows. Firstly, for the fixed coupling strength, the firing (depolarization-block) threshold, $I^n_1$ ($I^n_2$), is determined by the maximum (minimum) effective stimulus, $I^e_{max}=\max\{I^e_i, i=1,\ldots,N\}$ ($I^e_{min}=\min\{I^e_i, i=1,\ldots,N\}$). For the simplified model, we have $I^e_{max}=I^e_1$ and $I^e_{min}=I^e_5$. When $I$ is small, $I^e_i<I^s_1$ for all neurons and the network is at resting. As $I$ increases, $\{I^e_i\}$ will be increased altogether without changing their order [Fig. 4(a)], and, once $I^e_{1}$ exceeds $I^s_1$, neuron 1 will be fired. As a consequence, the network is transited from the resting to (partially) firing states. Increasing $I$ further, $I^e_i$ will be entering and then escaping from the firing region in sequence, and, once $I^e_5$ crosses $I^s_2$, the network will be transited from the firing to resting states. As such, we have $I^n_1\propto 1/I^e_1$ and $I^n_2 \propto 1/I^e_5$. Secondly, with the increase of $\varepsilon$, $I^e_1$ ($I^e_5$) will be decreased monotonically (increased) [as depicted in Fig. 4(f)]. Now, to make $I^e_1$ ($I^e_5)$ cross $I^s_1$ ($I^s_2$), a larger (smaller) $I$ therefore will be needed, resulting in the increased (decreased) $I^n_1$ ($I^n_2$). This explains the behavior of the firing region under strong couplings in the bifurcation diagrams. Thirdly, when the couplings are weak, the effective stimuli are widely spread [as shown in Fig. 4(g)]. In particular, the gap between $I^e_1$ and $I^e_2$ could be wider than the firing range $(I^s_1,I^s_2)$. In such a case, with the increase of $I$, there would be the situation where $I^e_1>I^s_2$ while $I^e_{2,3,4,5}<I^s_1$. As a result, a resting ``tongue", started from the point $I^e_1=I^s_2$ and ended at the point $I^e_2=I^s_1$,  will be appeared inside the firing region. This explains why under weak couplings the firing region is made up of disconnected subregions in the bifurcation diagrams [Figs. 2 and 3(b)].

If more than one large gaps (of width larger than the firing range) exist in the spectrum of the effective stimuli, more resting ``tongues" will be observed. For instance, when the coupling is weak (e.g., $\varepsilon=1.5$), the gap between $I^e_4$ and $I^e_5$ in the simplified model will be also wider than the firing range, resulting in the $2$nd resting ``tongue" in the bifurcation diagram [see the lower part of Fig. 3(b)]. This understanding is confirmed by numerical simulations, which show that in the left (middle, right) subregion, only neuron 1 [(2,3,4), 5] is fired. Based on this understanding, we can also predict that in the bottom-left corner of Fig. 2, there will exist some unique states where only the largest-degree neuron is fired. By scanning this corner with the improved precision, these special states are indeed observed, e.g., when $I=210~pA$ and $\varepsilon=0.1$, only neuron $1$ is fired in the network.

\section{Discussions and conclusion}

\begin{figure}[tbp]
\begin{center}
\includegraphics[width=0.85\linewidth]{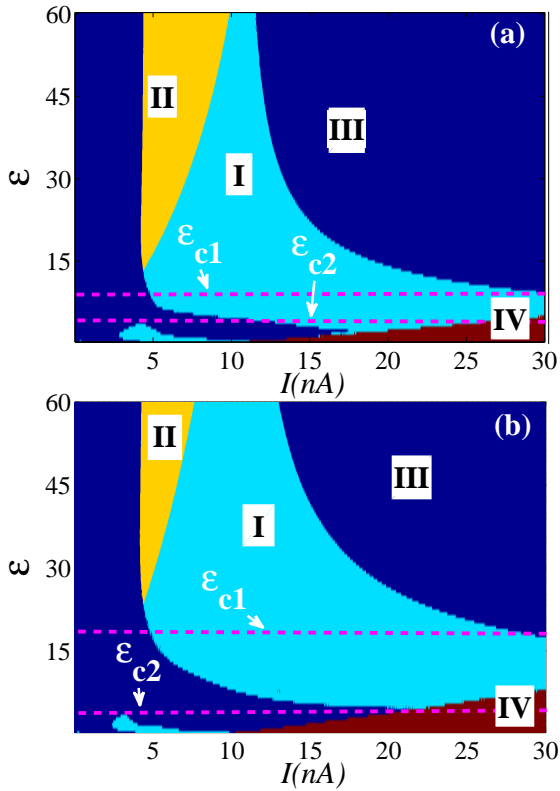}
\caption{(Color online) Bifurcation diagrams for the ER (a) and SW (b) networks. The size and the number of links of the networks are the same to that of BA network used in Fig. 2. Regions I, II, III, and IV represent, respectively, the partial-firing, complete-firing, resting, and overflowing states. For the ER (SW) network, the firing region is gradually narrowed as $\varepsilon$ increases from $\varepsilon_{c1}\approx 9.5$ ($\varepsilon_{c1}\approx 16.9$), and is separated into disconnected subregions in the region $\varepsilon<\varepsilon_{c2}\approx 4.2$ ($\varepsilon<\varepsilon_{c2}\approx 5.7$). The (magenta) dotted lines denote the critical coupling strengths $\varepsilon_{c1}$ and $\varepsilon_{c2}$.}
\label{Fig.5}
\end{center}
\end{figure}

Similar phenomena have been also observed for other network structures. Fig. 5(a) shows the bifurcation diagram for the Erd\"{o}s-R\'{e}nyi (ER) random network \cite{Net:ER}, which is constructed by connecting the existing neurons with an equal probability. The size and the number of links of the ER network are identical to that of the BA network used in Fig. 2, but the degrees are more homogeneously distributed. Still, the external stimulus is only injected on the largest-degree neuron (of degree $k_{max}=9$). Similar to Figs. 2 and 3(b), it is observed in Fig. 5(a) that in the region of strong couplings ($\varepsilon>\varepsilon_{c1}\approx 9.5$), the firing region is gradually enlarged as $\varepsilon$ decreases; while in the region of weak couplings ($\varepsilon<\varepsilon_{c2}\approx 4.2$) the firing region is separated into two disconnected subregions. Fig. 5(b) shows the bifurcation diagram for the small-world (SW) network \cite{Watts 1998}, which is constructed by rewiring randomly the connections of a regular network with the probability $p=0.1$. Again, it is seen that in the strong-coupling region ($\varepsilon>\varepsilon_{c1}\approx 16.9$) the firing region is gradually enlarged as $\varepsilon$ decreases, while in the weak-coupling regime ($\varepsilon<\varepsilon_{c2}\approx 5.7$) the firing region is separated into two disconnected subregions.

The present work is inspired by the series of experimental results reported in Refs. \cite{Brecht 2004,Houweling 2008,Morgan 2008,C. Y. Li 2009,Bi 2005}, where it is observed that the firing behaviors of the somatosensory cortex \cite{Brecht 2004,Houweling 2008}, or even the global brain \cite{Morgan 2008,C. Y. Li 2009}, could be significantly modulated by stimulating a single or a few neurons in vivo. For illustration purpose, we have chosen to stimulate only the largest-degree neuron in the network, yet it should be noted that the phenomena (and the firing mechanisms) we have observed (revealed) are general and independent of the stimulating location. For instance, the main features of the bifurcation diagram presented in Fig. 2 are still kept if the stimulus is injected on a neuron of medium degree ($k=8$). It is also worth mentioning that in the realistic brain network, a functional area is normally stimulated by a number of inputs at different locations, e.g., the rhythm signals received from the suprachiasmatic nucleus (SCN) \cite{RHYTHM}. In such a case, the network excitability will be also dependent of the stimulating strategy, i.e., the spatial configuration of the stimulus, and the bifurcation diagram might be significantly changed by adopting different stimulating strategies. A detail study to this question is out of the scope of the present work, which, hopefully, would be addressed elsewhere.

To summarize, we have studied the dynamical responses of complex neuronal networks subjected to a single stimulus, and found that the network could be evoked when the stimulating intensity is within a bounded region in the parameter space, namely the firing region. Furthermore, it is found that as the neuron coupling strength varies, the firing region is gradually modulated and, at the weak couplings, the firing region might be separated into disconnected subregions. By a simplified network model, we have conducted a detail analysis on the network firing mechanisms, and found the different roles that the stimulus and neuron couplings played in evoking the network. The findings shed new lights on the firing activities of complex neuronal networks, and might helpful in understanding some of the experimental observations \cite{Brecht 2004,Bi 2005,Houweling 2008,Morgan 2008,C. Y. Li 2009}

This work was supported by the National Natural Science Foundation of China under the Grant No.~11375109, and by the Fundamental Research Funds for the Central Universities under the Grant Nos.~GK201601001 and GK201503027.

\end{document}